\input{epsfig.sty}
\documentstyle[aps]{revtex}
\newcommand{\beq}{\begin{eqnarray}}
\newcommand{\eeq}{\end{eqnarray}}
\begin{document}
\title{The $\Lambda$(1600): A Strange Hybrid Baryon}
\author{Leonard S. Kisslinger$^\dagger$\\
    Department of Physics, Carnegie Mellon University, Pittsburgh, PA 15213}
\maketitle
\indent
\begin{abstract}
  We use the method of QCD sum rules to investigate a possible hybrid
baryon with the quantum numbers of the $\Lambda$. Using a current composed
of uds quarks in a color octet and a gluon, a strange hybrid, the $\Lambda_H$,
is found about 500 MeV above the $\Lambda$, and we identify it as the 
$\Lambda(1600)$  Using our sigma/glueball model we predict a large branching 
fraction for the $\Lambda_H \rightarrow \Lambda+\sigma(\pi\pi$
resonance), and the experimental search for this decay mode could provide
a test of the hybrid nature of the $\Lambda(1600)$.
\end{abstract}

\vspace{5mm}

\noindent
PACS Indices:12.39.Mk,14.20.Jn,12.38.Lg,11.55.Hx

\vspace{2 mm}
$^\dagger$ email: kissling@andrew.cmu.edu

\section{Introduction}

   One of the most important aspects of hadron spectroscopy for providing
new information about Quantum Chromodynamics (QCD) is the identification
of hybrid hadrons, mesons and baryons that have gluonic valence components.
These are typical exotics, hadrons that cannot be explained in standard
quark models.  Since some exotic mesons can be identified by their quantum
numbers,  great deal of work has been done on hybrid mesons. Hybrid baryons
are more difficult to identify experimantally, although there are clear
theoretical definitions. For example, a nonstrange baryon or a baryon with 
strangeness = -1 with three-quark components in
a configuration with color $\neq$ 0, while the three quarks plus gluon
have zero color is certainly a hybrid. Strangeness = +1 baryons are exotic,
but cannot be characterized as hybrids in the sense that they must involve
an anti-strange quark rather than valence glue.

There have been theoretical studies 
of hybrid baryons for two decades, including bag model  of nonstrange hybrids
calculations\cite{bc83,gol83,duck83} and of strange hybrids\cite{carl83},
models of the $P_{11}(1440)$ (Roper) resonance as a hybrid\cite{zl,lbl}, 
QCD Sum Rule studies\cite{mar91,kl95,k98} and flux tube models\cite{cp99,cp02}.
A major problem is to identify the hybrids. Electroproduction\cite{lbl} and
large decay branching ratio into the sigma $\pi-\pi$ resonance\cite{kl99}
have been suggested for tests of the Roper resonance as a hybrid;
and possible tests of hybrids related to the flux tube model have been 
suggested\cite{page00}. See the recent review by Barnes\cite{barnes00} for 
a discussion of hybrid baryons and possible ways to identify them.

  In the early bag model studies of nonstrange hybrid baryons
\cite{bc83,gol83,duck83} there was a range of solutions of about 1.5-1.9 GeV 
for the lightest hybrid. Using a bag model for the study of $\Lambda$-like 
strange hybrid baryons, i.e., using (udsg) configurations with the uds quarks 
in a color octet, in Ref.~\cite{carl83} a narrower range of solutions was 
found with the choice of parameters used, and the authors concluded the the 
lightest $\Lambda$-type hybrid might be the three-star $\Lambda(1600)$.

    In the present work we use the QCD sum rule method as in Ref.~\cite{kl95}
to estimate the mass of the lightest strange baryon with the quantum numbers
of the $\Lambda: J^P = \frac{1}{2}^+,  S=-1, I=0$. One advantage of the
sum rule method is that the composite field operator, called the current,
for the particle, for which we use the symbol $\Lambda_H$, is a clearly 
defined hybrid, with the three quarks having nonzero color. Our conclusion
in the present work is that the $\lambda(1600)$ is a hybrid. The theoretical 
calculations is similar to that of our previous work in which we concluded 
that the Roper is a good hybrid baryon candidate. Note that the excitation 
energy of the  $P_{11}(1440)$ resonance above the nucleon is similar to that 
of the $\Lambda(1600)$ above the $\Lambda$. We also use 
the arguments of Ref.~\cite{kl99} to suggest a test of the  $\Lambda(1600)$
as a hybrid.

 \section{QCD Sum Rule Estimate of the mass of the $\Lambda_H$}

    The current operator for the $\Lambda_H$ which we use is the form used in
Ref.~\cite{kl95}
\beq
\label{etaH}
   \eta_{\Lambda H} &=& \epsilon^{abc}[u^a(x)C\gamma^\mu d^b(x)]\gamma^\alpha
 G^e_{\mu\alpha}(x) [\frac{\lambda^e}{2}s(x)]^c,
\eeq
where (a,b,c,e) are color indices, $\lambda^e$ is the generator of the SU(3)
color group, (u,d,s) refer to up, down and strange quarks, and
$G_{\mu\nu} = \partial_\mu A_\nu - \partial_\nu A_\mu-i g [A_\mu,A_\nu]$ is
the gluonic tensor, with $A_\mu = A_\mu^n \lambda^n/2$  the color field.
The corresponding correlator is

\beq
\label{correlator}
 \Pi_{\lambda H}(x) &=& <0|T[\eta_{\Lambda H}(x)\bar{\eta}_{\Lambda H}(0)]|0> 
\nonumber \\
          &=& \epsilon^{abc}\epsilon^{a'b'c'}\gamma_5\gamma_\alpha S_s^{ee'}(x)
\gamma_\beta \gamma_5 <0| G^m_{\mu\alpha}(x) G^n_{\nu\beta}(0)|0>
 \frac{\lambda^m_{ce}}{2}\frac{\lambda^n_{c'e'}}{2}
\nonumber \\
  && Tr[S^{aa'*}_u(x)C\gamma^\mu S^{bb'}_d(x)\gamma^\nu C],
\eeq
where $S_f$ is the flavor f quark propagator and $C$ is te charge conjugation
operator. The correlator has two independent terms, which in momentum space can
be written as
\beq
\label{qcorrelator}
   \Pi(q)_{\Lambda H} &=& i\int e^{iqx}  <0|T[\eta(x)\bar{\eta}(0)]|0>
 \nonumber \\
          &=& \Pi_1(q^2)\not\!q + \Pi_2(q^2),
\eeq
with $\not\!q = \gamma_\mu q^\mu$.

\begin{figure}
\begin{center}
\epsfig{file=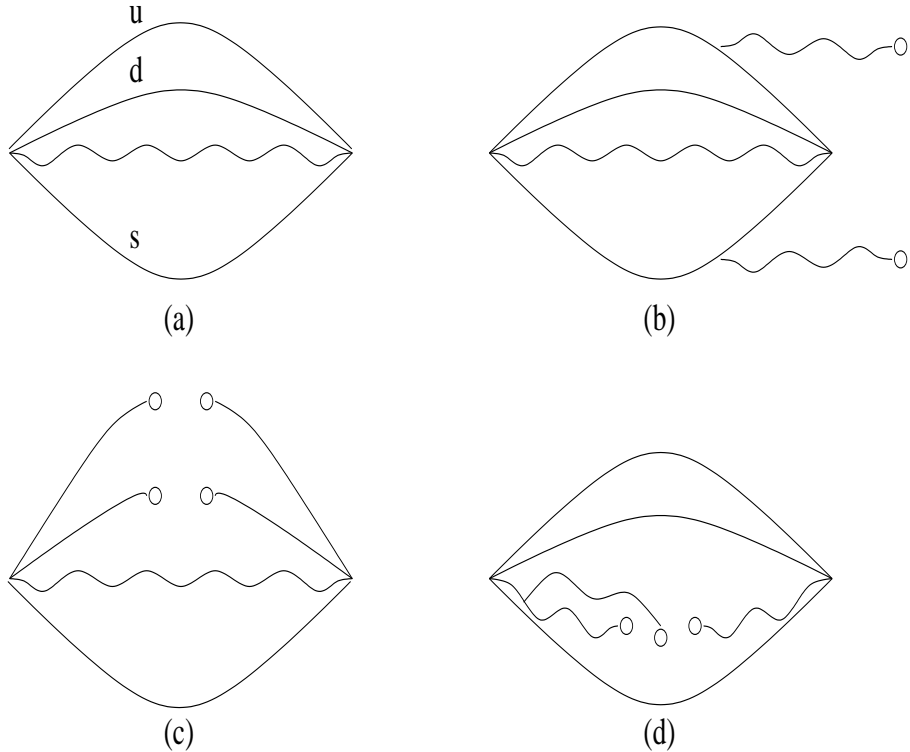,height=10cm,width=12cm}
\caption{Diagrams for $\Pi_1(q^2)$ for the OPE including dimension 6 
condensates}
{\label{Fig.1}}
\end{center}
\end{figure}

\begin{figure}
\begin{center}
\epsfig{file=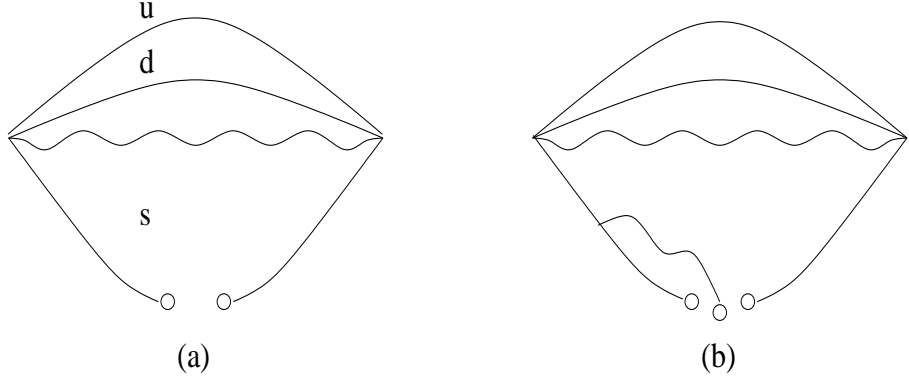,height=5cm,width=12cm}
\caption{Diagrams for $\Pi_2(q^2)$ for the OPE including dimension 5 
condensates}
{\label{Fig.2}}
\end{center}
\end{figure}

   The QCD sum rule calculation is very similar to that in 
Refs.~\cite{kl95,k98}. The operator product expansion converges rapidly
at high $Q^2 = -q^2$, which is acheived via a Borel transform. The diagrams
included for $\Pi_1(q^2)$ are shown in Fig.~\ref{Fig.1}
and the diagrams included for $\Pi_2(q^2)$ are shown in Fig.~\ref{Fig.2}

   One finds from the diagrams of Fig.~\ref{Fig.1} that
\beq
\label{Pi1}
   \Pi_1(Q^2) &=& \frac{1}{45 \pi^6} Q^8ln(Q^2) -\frac{17}{36 \pi^5}
 <0|\alpha_s G^2|0> Q^4 ln(Q^2) \nonumber \\
 && +\big( \frac{512}{9 \pi^2} <0|\bar{q}q|0>^2
- \frac{2}{3 \alpha_s \pi^5} <0|g^3 fG^3|0> \big)Q^2 ln(Q^2),
\eeq
and from the diagrams of Fig.~\ref{Fig.2} that
\beq
\label{Pi2}
   \Pi_2(Q^2) &=& \frac{16}{9 \pi^4} <0|\bar{s}s|0> Q^6 ln(Q^2)
 -\frac{20}{\pi^4}<0|\bar{q}\sigma \cdot G q|0> Q^4 ln(Q^2),
\eeq
where $Q^2=-q^2$, and for the condensates the values used are:
quark condensate $<0|\bar{q}q|0> = -(.25)^3$ GeV$^3$, mixed condensate
$<0|\bar{q}\sigma \cdot G q|0> = 0.8 (.25^3)$ GeV$^5$, the gluon condensate
$<0|\alpha_s G^2|0> = 0.038$ GeV$^4$, and the triple gluon condensate
$<0|g^3 fG^3|0> = (0.6)^6$ GeV$^6$. The convention of Ref.\cite{mar91}
has been used. In deriving Eqs.(\ref{Pi1},\ref{Pi2}) the quark masses have been
neglected. Also the strange quark condensate $<0|\bar{s}s|0>$ is expected
to be about 20\% smaller than the u,d quark condensates, but this difference 
is neglected as it does not change our results significantly.

   The sum rule method involves equating the dispersion relation for
the correlator to the form obtained from the OPE expansion, using a Borel
transfrorm to ensure that the OPE converges rapidly and that the
continuum contribution to the dispersion relation is not to large to
find the mass given by the pole term. For our problem we must take into
account that the $\Lambda$ as well as the $\Lambda_H$ can contribute to
the correlator given by Eqs.(\ref{etaH},\ref{correlator}), since the
$\Lambda$ can have a (udsg) component. Thus the form of the dispersion
relation for the correlator, called the phenomenological side of the sum 
rule, is
\beq
\label{phen}
      \Pi_1^{phen}(Q^2) &=& \frac{c_1}{Q^2 + M_\Lambda^2} +
 \frac{c_2}{Q^2 + M_{\Lambda H}^2} + {\rm continuum} \nonumber \\
       \Pi_2^{phen}(Q^2) &=& \frac{c_1 M_\Lambda}{Q^2 + M_\Lambda^2} +
 \frac{c_2 M_{\Lambda H}}{Q^2 + M_{\Lambda H}^2} +{\rm continuum}.
\eeq
After the Borel transform, in which the $Q^2$ variable is replaced by
Borel mass $M^2$, the phenomenological side becomes
\beq
\label{phenM}
    \Pi_1^{phen}(M^2) &=& c_1 e^{-M_{\Lambda}^2/M^2} 
 +c_2 e^{-M_{\Lambda H}^2/M^2} + {\rm continuum} \nonumber \\
   \Pi_2^{phen}(M^2) &=&  c_1 M_\Lambda e^{-M_{\Lambda}^2/M^2} +
c_2 M_{\Lambda H} e^{-M_{\Lambda H}^2/M^2} + {\rm continuum} \; .
\eeq
The constants $c_1, c_2$ in Eq.(\ref{phen}) contain important information
about the wave function, but can be eliminated in the analysis by using the 
technique of Ref.\cite{kl95}. The continuum is approximated by replacing the
Borel transform
\beq
\label{borel}
    {\cal B}_{Q^2 \rightarrow M^2}(Q^{2l}lnQ^2)= -l!M^{2(l+1)}
\eeq
by
\beq
\label{borelc}
   {\cal B}^c_{Q^2 \rightarrow M^2}(Q^{2l}lnQ^2)= -M^{2(l+1)}
  \big(1-\sum_{n=1}^{l}\frac{(s_t/M^2)^n}{n!}e^{-s_t/M^2} \big),
\eeq
where $s_t$ is the threshold $E_{c.m.}^2$, which is chosen by
matching the phenomenological to the theoretical expressions for the
correlator.

   The mass of the hybrid is obtained from the expression
\beq
\label{mass}
   M_{\Lambda H} &=& \frac{\frac{d}{dM^2}[\Pi_2(M^2)e^{M_\Lambda^2/M^2}]}
 {\frac{d}{dM^2}[\Pi_1(M^2)e^{M_\Lambda^2/M^2}]}.
\eeq
The sum rules for $\Pi_1$ and  $\Pi_2$ are quite stable, and the solution
give a hybrid mass about 500 MeV above the $\lambda$ mass. An error of
about 15\% is expected with this method. There we conclude that 
\beq
\label{massH}
   M_{\Lambda H} & \simeq & 1600 \pm 200 \; {\rm MeV},
\eeq
and that the $\Lambda(1600)$ is our candidate for the lightest
strange hybrid, with the quantum numbers of the $\Lambda(1115)$.

 \section{The $\sigma$ Decay of the $\Lambda_H$}

   A test of the hybrid nature of the $\Lambda(1600)$ follows from our
glueball/sigma model\cite{k97,k01}. In studies of the scalar hadrons, 
the mixing of scalar mesons and scalar glueballs was found to be 
important for mesons and glueballs with masses above 1 GeV. Moreover,
a light glueball solution was obtained with a mass 300-600 MeV, which
is the mass of the broad I=0, L=0 $\pi-\pi$ resonance, which was found in 
an analysis of $\pi-\pi$ scattering\cite{zb95}, which we call the sigma. 
This leads to our glueball/sigma model, a model based on a coupled-channel 
picture with the glueball pole driving the $\pi-\pi$ resonance, motivated in 
part by the large branching ratios for decays of glueball candidates into 
channels with sigmas\cite{bes96}. This model successfully 
explains\cite{lsk01} the observed\cite{E791} $D^+$ charm meson decay into 
a $\sigma (\pi\pi)$ resonance, while this decay channel is not found for 
the $D_s^+$ decay.

\begin{figure}
\begin{center}
\epsfig{file=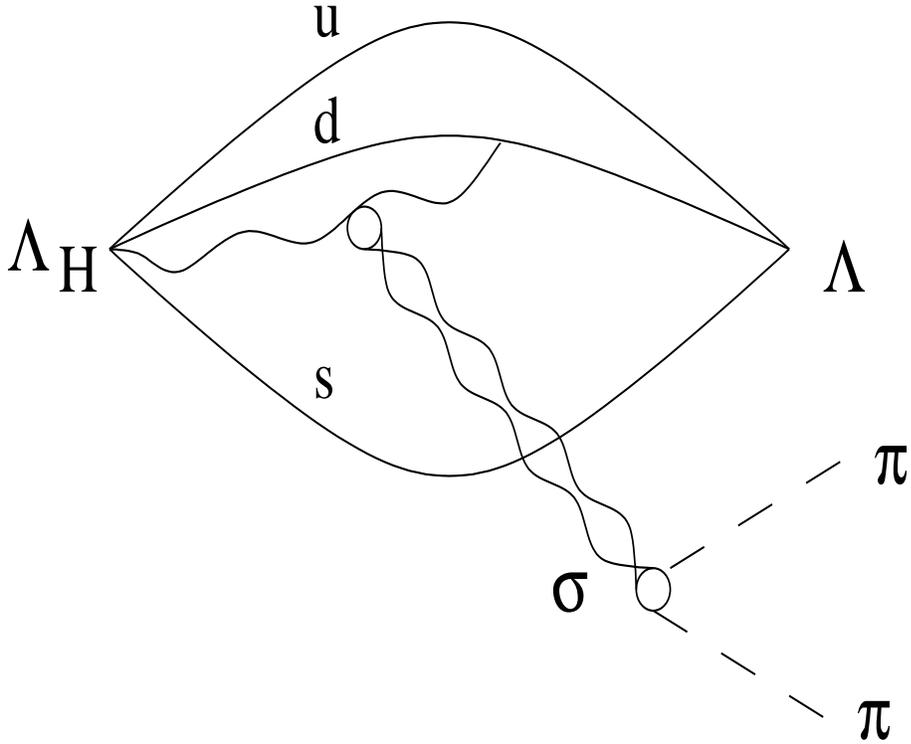,height=10cm,width=12cm}
\caption{Diagram for $\Lambda_h \rightarrow \Lambda + 2\pi$ via qlue/
$\sigma$ coupling}
{\label{Fig.3}}
\end{center}
\end{figure}

   Using the current operator for the $\Lambda$
\beq
\label{etaL}
   \eta_{\Lambda} &=& \epsilon^{abc}[u^a(x)C\gamma^\mu d^b(x)]
 \gamma_5 \gamma_\mu s(x)^c \; ,
\eeq
the decay of the $\Lambda_H$ into a $\Lambda$ and a sigma can be
estimated by the external field method\cite{is84} from the 
$\Lambda_H,\Lambda$ two-point function in an external Sigma field 
\beq  
\label{external}
   \Pi(q)_{\Lambda \Lambda_H} &=& i\int e^{iqx}  <0|T[\eta_\Lambda(x)
\bar{\eta}_{\Lambda H}(0)]_{J_\sigma}|0>,
\eeq
where $J_\sigma$ is the sigma current. E.g., if one were investigating
the decay into the pion channel, which is a forbidden $\Lambda_H
\rightarrow \Lambda$ decay, the current would be $J^\pi = i\bar{q}
\gamma_5 q$. The calculation is illustrated in Fig.~\ref{Fig.3} 
for the lowest dimensional diagram.

   The glueball sigma model leads to the evaluation of the sigma-gluonic
coupling
\beq
\label{gsig}
  <G^a_{\mu\nu} J_\sigma G^{\mu\nu}_a> &=& g_\sigma<G^a_{\mu\nu} G^{\mu\nu}_a>.
\eeq
The sigma-glue coupling constant, $g_\sigma$, was estimated from the
356 Mev width of the $\sigma(\pi\pi)$ resonance\cite{zb95} to be $g_\sigma 
\simeq$ 700 MeV. Because of the unknown constants in the sum rule
calculation one can only predict ratios of decay widths. For the study
of the Roper as a hybrid we were able to predict the ratio of the sigma 
decay width to the pion decay width. Since the only decay channels that have 
been measured for the $\Lambda(1600)$  are
$\Lambda(1600) \rightarrow N \bar{K}$ and $\Lambda(1600) \rightarrow
\Sigma \pi$, and the single-pion + $\Lambda$ channel is not allowed,
we are not able to make a specific prediction here. Noting 
that the decay $P_{11}(1440) \rightarrow N + (\pi\pi)_{I=0,L=0}$ has
a 5-10 \% branching fraction, we predict a sizable $\Lambda_H \rightarrow
\Lambda + \sigma$ branching fraction. The main background will be the
nonresonant $\pi-\pi$ channels, so that for the sigma channel
the (I=0,L=0) resonance must be extracted. Also, it would be
difficult to measure the ratio $\Lambda_H \rightarrow \Sigma \sigma/
\Lambda_H \rightarrow \Sigma \pi$, for which our model could give a
prediction, since there is very little of phase
space to map out to extract the $\sigma$ resonance due to the mass of
the $\Sigma$.
\vspace{3mm}

   In conclusion, I believe that a careful experimental study of the  
$\Lambda(1600) \rightarrow \Lambda \pi \pi$, with an analysis to extract
the sigma resonance, could be a valuable test of the hybrid nature of
this resonance.  There have been no experiments on $\Lambda(1600)$ decays
for past twenty years, and this resonance is an excellent candidate for
a strange hybrid baryon.
\vspace{3mm}

   This work was supported in part by NSF grant PHY-00070888 and in part by
the DOE contract W-7405-ENG-36. The author thanks the P25 group at LANL
for hospitality while part of this work was being carried out, and
Dr. Reinhard Schumacher for valuable discussions.

\end{document}